\documentclass{pseudoelsart} 
\usepackage{latexsym} 
\usepackage{alltt}
\usepackage[all]{xy} 
\CompileMatrices 
\xyoption{frame}
\usepackage{amssymb}
\usepackage{amscd} 
\usepackage{amsmath}
\usepackage{theorem} 
{\theorembodyfont{\rmfamily}
} 
{\theorembodyfont{\rmfamily}
} 
{\theorembodyfont{\rmfamily}
}
{\theorembodyfont{\rmfamily}
}

\newcommand{\tr}{{\mathtt{true}}}
\newcommand{\fa}{{\mathtt{false}}}

\newcommand{\prom}[3]{#1 \overset{\pi:#2}{- \!\!\!- \!\!\!- \!\!\!- \!\!\!\!\!\!\!\!\longrightarrow} #3}
\newcommand{\lprom}[3]{pi_{#1,#2}(#3)}
\newcommand{\lwprom}[3]{pw_{#1,#2}(#3)}
\newcommand{\geb}{{\sim}}

\begin{document} 
\begin{frontmatter} 
\title{A process algebra based framework for promise theory}
\author[label1,label2]{Jan  Bergstra\thanksref{email1}} 
\author[label1]{Inge Bethke\thanksref{email2}\corauthref{cor}}
\author[label3]{Mark Burgess\thanksref{email3}}
\corauth[cor]{Corresponding author. Address: Kruislaan 403, 1098 SJ Amsterdam, The Netherlands }
\thanks[email1]{E-mail: \texttt{janb@}$\{$\texttt{phil.uu.nl, science.uva.nl}$\}$}
\thanks[email2]{E-mail:  \texttt{inge@science.uva.nl}}
\thanks[email3]{E-mail:  \texttt{mark@iu.hio.no}\\
\mbox{}\\
An earlier version 
of this paper also appeared as report PRG0701, Section Theoretical Software Engineering, 
Informatics Institute, 
Faculty of Science, University of Amsterdam.}
\address[label1]{University of Amsterdam, Faculty of Science, Section Theoretical
Software
Engineering (former Programming Research Group)} 
\address[label2]{Utrecht University, Department of Philosophy, Applied Logic
Group}
\address[label3]{University College Oslo, Faculty of Engineering}
\begin{abstract}
We present a process algebra based approach to formalize the interactions of computing devices
such as the representation of policies and the resolution of conflicts. As an example we specify how
promises may be used in coming to an agreement regarding a simple though
practical transportation problem.
\end{abstract}
\end{frontmatter}
\section{Introduction}
The mechanism of an autonomous agent announcing a promise towards another agent
is a powerful organizational principle in the setting of computer networks.
Several approaches have been used in the past to formalize such interactions of
computing devices as a
representation of policies and a resolve of conflicts: 
Burgess and Fagernes \cite{BF06}
represent them as graphs, Prakken and Sergot \cite{PS96,PS97} use temporal
deontic logic, Lupu and Sloman \cite{LS97} propose role theory, Glasgow et al.
\cite{GMP92} modal logic, Bandera et al. \cite{BLMR04} event
calculus and  Lafuente and Montanari \cite{LM05} model checking.
In this paper we use process algebra \cite{BPS01} for the formalization of a
restricted set of aspects of promises paying attention to the sequential
ordering of promises between a number of parties. As an example we specify how
promises may be used in coming to an agreement regarding a simple though
practical transportation problem.

In the world of process algebra we can label certain communications as promises
if that makes sense intuitively. Process algebra formalisms will not provide very
sharp distinctions that set apart \emph{promise acts} from all other conceivable
actions, however. Modal logics are in principle better suited for the task to
capture what is specific about promises, but process algebras may be more
helpful to formalize the role that  promises can play in specific multi-agent
systems. The justification of the
process algebra framework for promises is therefore as follows:
\begin{enumerate}
\item to provide clear and formalized cases of the use of promises in some
protocols that occur within multi-agent systems,
\item to support the design and analysis of distributed protocols that make use
of promises made by autonomous agents.
\end{enumerate}
The process algebra framework cannot, by nature, characterize the concept of a
promise in its logical essence. That is a much harder task and requires the
design of specific versions of deontic logic.

\section{A data type for task bodies}
\begin{figure}[htbp] 
\centering
\fbox{ 
\mbox{ 
\xy 
\UseComputerModernTips 
\xymatrix@=40pt@!{ 
&{\mathcal{T}}\\
{\mathcal{TB}} \ar@(l,u)[]^{\geb} \ar@(l,d)[]_{\neg}\ar[ur]^{t}
\ar@{>->}[dr]|{\#} \ar@/^1pc/[dr]^p \ar@/_1pc/[dr]_{s}&\ar[l]_\gamma\\
&{\mathbb{B}}\ar@(r,u)[]_\neg
\\
&\ar@/^1pc/[u]^\tr \ar@/_1pc/[u]_\fa\\
} 
\endxy 
}} 
\caption{Data type for task bodies}\label{TB} 
\end{figure}
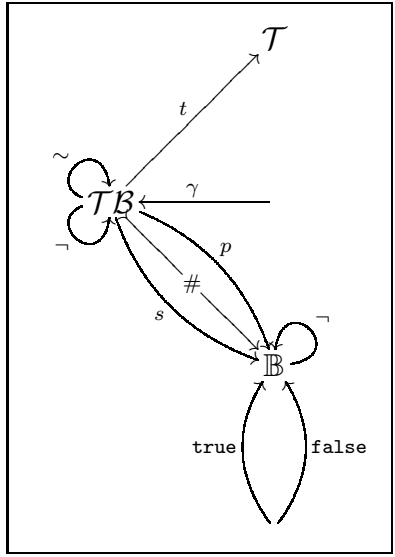

The data type for task bodies is depicted in  Figure \ref{TB}.

$\mathcal{TB}$ is assumed to be a finite set of primitives which fall into two
basic complementary
categories, namely into tasks for giving or taking, or \emph{services} and \emph{usage}.
We distinguish the atom $\gamma$|the special task of compliance.
Atomic tasks are assumed services rather than uses and positive, i.e.\ , not 
negated . Two operations $\geb  , \neg : \mathcal{TB} \rightarrow \mathcal{TB}$ on 
tasks
are then considered:
\begin{enumerate}
\item (\emph{usage}) if $x$ is a task then $\geb  x$\footnote{In \cite{B06,BF06}
the use of a $x$ is denoted by $-x$ or $U(x)$ instead of $\geb x$.} is the task of making use of $x$ as performed by
another agent, and
\item (\emph{negation}) if $x$ is a task then $\neg x$ is the task of not doing $x$.
\end{enumerate}
Moreover, $s, p: \mathcal{TB} \rightarrow \mathbb{B}$ specify the properties \emph{service} and 
\emph{positive}.
The interaction of these operations satisfies the laws in Figure \ref{usage_negation}.
\begin{figure}[htbp]
\[
\begin{array}{|crclc|crclc|crclc|}
\hline
&&&&&&&&&&&&&&\\
\hspace{0.5cm}&\geb  \geb  x & = & x&\hspace{0.5cm}&\hspace{0.5cm}&
s(\gamma) 
& = & \tr&\hspace{0.5cm}&\hspace{0.5cm}&p(\gamma)&= &\tr&\\[2mm]
&\neg \neg x & = & x&&&s(\neg x) & =& s(x)&&& p(\neg x)&=& \neg p(x)&\\[2mm]
&\geb  \neg x & = & \neg \geb   x&&&s(\geb x)&=& \neg s(x)&&&p(\geb x)&=& p(x)&\\[2mm]
\hspace{.5cm}&&&&&&&&&&&&&&\hspace{.5cm}\\
\hline
\end{array}
\]
\caption{Interaction of usage, negation and service \label{usage_negation}}
\end{figure}
Note that $\neg$ is overloaded in the sense that it acts as negation on tasks and
on Booleans. The actual meaning, however, will always be clear from the context.

In general, promises can be viewed as declarations to keep certain tuples of data 
within a given range of values. Promises are thus typed. We therefore assume a collection of types, $\mathcal{T}$,
and a typing function $t: \mathcal{TB} \rightarrow \mathcal{T}$ providing types
for task bodies.  Given a service $x$, we assume
that types do not differ under usage or negation, i.e.,
\[
t(\geb  x)=t(x)=t(\neg x).
\]
Furthermore, since promises can be incompatible with each other we assume  a symmetric
incompatibility relation $\#: \mathcal{TB}\times \mathcal{TB} \rightarrow  \mathbb{B}$.
We write $x \# y$|instead of $\#(x,y) = \tr$|if $x$ and $y$ \emph{cannot} both be realized
at the same time by the same agent. 
Only tasks of similar type can exclude one another. Moreover, 
tasks are incompatible with their negations. 
For incompatible tasks $x$ and $y$, however, $x$ will be compatible with $\neg y$.
\begin{figure}[htbp]
\centering
\fbox{
\begin{minipage}{10cm}
\mbox{}\\
\[
x \# \neg x\ \ \ \ \ \frac{x\# y}{y \# x}
\ \ \ \ \ \frac{x \# y}{t(x)=t(y)}
\ \ \ \ \ \ \frac{x\# y}{\neg ( x \# \neg y)}
\]
\mbox{}\\
\mbox{}
\end{minipage}
}
\caption{Laws of incompatibility}\label{incom}
\end{figure}
Observe that 
\(
\neg (x \# x)
\)
is derivable from the axiom and the third rule in Figure \ref{incom} using the law of double negation shift.

\section{A transition system for promises}
Let $A$ be a partially ordered set containing so-called agents. For agents $a,b\in A$, we write $a\leq b$ if $a$ is subordinated to $b$.
We denote a promise $x$ between arbitrary autonomous agents $a$ and $b$|while being unspecific about how and 
when they are made|by \[\prom{a}{x}{b}.\]
For $x\in \mathcal{TB}$ with $s(x)=\tr=p(x)$ we distinguish the 4 kinds of promises
given in Figure \ref{4prom}
\begin{figure}[htbp]
\[
\begin{array}{|cll|}
\hline&&\\
\hspace{0.5cm}&(1)\hspace{1 cm}&\prom{a}{x}{b}\hspace{1.5cm}\\[4mm]
&(2)&\prom{a}{\geb  x}{b}\\[4mm]
&(3)&\prom{a}{\neg x}{b}\\[4mm]
&(4)&\prom{a}{\neg \geb   x}{b}\\[4mm]
\hline
\end{array}
\]
\caption{Promised exchanges of services $x$ between autonomous agents $a$ and 
$b$\label{4prom}}
\end{figure}
where
\begin{enumerate}
\item $a$ promises $b$ to provide its service $x$,
\item $a$ promises $b$ to make use of its service $x$,
\item $a$ promises $b$ not to provide its service $x$, and
\item $a$ promises $b$ not to make use of its service $x$.
\end{enumerate}
We tacitly assume that promises are equal under equal tasks, i.e., that
\[
x=y \Rightarrow \prom{a}{x}{b}=\prom{a}{y}{b}.
\]

One can generalize this basic notation of promise exchange to a more expressive system where agents can make promises about what other agents
might do|provide a service or make use of. A generalized notation of the form
\[
\prom{a[c]}{x}{b[d]}
\]
denotes that `$a$ promises $b$ that $c$ will do $x$ for $d$'.
The autonomously made promises in Figure \ref{4prom} are then equivalent to their more general notations
\[
\prom{a[a]}{x}{b[b]}.
\]

If $c\leq a$ and $\prom{a[c]}{x}{b[d]}$, then the promise by $a$ implies an obligation for $c$. Autonomous agents, however, ought not to be obliged to anything.

Another interaction between autonomous promises and the more general kind of promises is given by the so-called
\emph{compliance promise} between agent $c$ and $a$,
\[
\prom{c}{\gamma}{a},
\]
where $c$ promises to comply with $a$. We then have
\[
\prom{a[c]}{x}{b[d]},\prom{c}{\gamma}{a} \Longrightarrow \prom{c}{x}{d}.
\]

One can consider the even more general notation\[
\prom{a[c_1,\ldots, c_n]}{x}{b[d_1,\ldots, d_m]}
\]
denoting that 
\begin{enumerate}
\item `$a$ promises $b$ that one of $c_1, \ldots, c_n$ wil do $x$ for some one amongst $d_1, \ldots, d_m$' if $x$ is a service, or
\item `$a$ promises $b$ that one of $c_1, \ldots, c_n$ will make use of $x$ as done by one amongst $d_1, \ldots, d_m$' if $x$ is a usage
\end{enumerate}
provided $x$ is positive. In the negative case none of $c_1, \ldots, c_n$ wil do $x$ for any of $d_1, \ldots, d_m$' etc.

In distributed systems design it is unhelpful to use either $\prom{a[c]}{x}{b}$ or $\prom{a[c]}{x}{b[d]}$. If these occur in a design they should
and usually can be translated into small protocols using \emph{voluntary} promises only. In the sequel we will therefore 
focus on promises of the basic form made by autonomous agents forgetting about the general notion of promises.

We will model states as sets of basic promises that do not conflict together with
transition rules that describe the development of such states. The presence of a single promise  \[\prom{a}{x}{b}\] is written as $p_{a, b}(x)$ and promises are combined by the promise set composition operator $\oplus$.

A \emph{transition rule} for a basic promise has one of the two forms
\[
pi_{a, b}(x)\ \frac{S}{S\oplus \prom{a}{x}{b}}\ \ \ \ \ \textit{promise introduction}
\]
or
\[
pw_{a, b}(x)\ \frac{S}{S\ominus \prom{a}{x}{b}}\ \ \ \ \ \textit{promise withdrawal}
\]
where
\begin{enumerate}
\item $S$ is a state, i.e., a set of non-conflicting basic promises,
\item the promise introduction $\lprom{a}{b}{x}$ labels the transition rule with
the announcement that introduces the promise
$\prom{a}{x}{b}$, 
\item the promise withdrawal $\lwprom{a}{b}{x}$ labels the transition rule with the speech act that withdraws the promise
$\prom{a}{x}{b}$, 
\item $\oplus$ combines the state $S$ with the promise $\prom{a}{x}{b}$ yielding a new state, and
\item $\ominus$ removes the promise $\prom{a}{x}{b}$ from the state $S$ yielding a new state.
\end{enumerate}
Since states are sets of non-conflicting promises, a promise introduction event (that is an application of the promise introduction rule) is applicable only if the conclusion of the rule is a set of non-conflicting
promises, i.e., if for all $\prom{a}{y}{c}\in S$, $\neg (x\#y$).
Here we assume that an autonomous agent is itself responsible
for making no promises that would require performing incompatible
tasks (`breaking its own promises' is Burgess' nomenclature in \cite{B06}).

This system can be generalized to generalized promises. A typical rule in this format is of the form
\[
pi_{a[c]\rightarrow b[d]}(x)\ \frac{S\oplus \prom{c}{\gamma}{a}}{S\oplus \prom{c}{\gamma}{a}\oplus \prom{c}{x}{d}}
\]

In addition to incompatibility we now introduce \emph{exclusiveness} $E: \mathcal{TB} \rightarrow \mathbb{B}$
marking tasks that cannot be served to or consumed from two different agents at the same time. This will mean that
$\prom{a}{x}{b}$ and $\prom{a}{x}{c}$ with $b\neq c$ can never be kept if $E(x)=\tr$|and should not 
both be made either. In the presence of exclusiveness, the promise event rule takes the conditional format
\[
(E(x) \rightarrow \forall c\neq b\ \neg p_{a,c}(x)) \Longrightarrow pi_{a, b}(x)\ \frac{S}{S\oplus \prom{a}{x}{b}},
\]
Note that exclusiveness is not related to incompatibility: `taking a train' and 
`taking a car'
are conflicting tasks; `being driven by' $b$, however, excludes `being driven by' $c$.

\section{An example}

We now consider an ACP-style process algebra with the standard
operators $+, \cdot, \parallel$ for choice, sequential and parallel composition (cf. \cite{BW90,F00}), and conditional guards (cf.\ e.g. \cite{BB92}) 
based on
atomic actions like $pi_{a,b}(x)$ and $pw_{a,b}(x)$. In such a setting a protocol, $P_{a,b}(x)$, that describes
a plausible course of actions for introducing a promise $\prom{a}{x}{b}$ can be given by
\[
\begin{array}{rl}
P_{a,b}(x) = pi_{a,b}(x) \cdot ( & (E(\geb x)\rightarrow \forall c\neq a\ \neg p_{b,c}(\geb x)):
\rightarrow pi_{b,a}(\geb x)\\
& + \\
&pi_{b,a}(\neg \geb x) \cdot (pw_{a,b}(x) \parallel pw_{b,a}(\neg \geb x))\\
)&
\end{array}
\]
Here is an example from our recent experience. The  autonomous agents 
Jan, J\"urgen, and Mark consider the task of transport by car to the Jacobs
University Bremen (JUB), i.e.,
\begin{enumerate}
\item $A=\{ja, ju, ma\}$, and
\item $\mathcal{TB}=\{\mathit{tbc2JUB}, \neg \mathit{tbc2JUB}, \geb \mathit{tbc2JUB}, \neg \geb \mathit{tbc2JUB}\}$.
\end{enumerate}
Since one cannot be transported in 2 different cars at the same time and by 2 
different people, $\geb \mathit{tbc2JUB}$ is exclusive, i.e.,
\[
E(\geb \mathit{tbc2JUB}). 
\]
A possible execution of $P_{ja,ma}(\mathit{tbc2JUB})\parallel P_{ju,ma}(
\mathit{tbc2JUB})$ is given by the trace
\[
\begin{array}{ll}
pi_{ja,ma}(\mathit{tbc2JUB})&\cdot\\
pi_{ma,ja}(\mathit{\geb tbc2JUB})&\cdot\\
pi_{ju,ma}(\mathit{tbc2JUB})&\cdot\\
pi_{ma,ju}(\mathit{\neg \geb tbc2JUB})&\cdot\\
pw_{ju,ma}(\mathit{tbc2JUB})&\cdot\\
pw_{ma,ju}(\mathit{\neg \geb tbc2JUB}).&\\
\end{array}
\]
On an intuitive level, the trace can be described as follows:
Initially Jan promises Mark a lift to JUB which Mark accepts. Then 
J\"urgen makes this promise too which Mark|because of the exclusiveness of this
task|declines. Thereupon J\"urgen withdraws his offer and Mark his declination.

This kind of example can typically be found in data centre management:
renaming the agents and tasks to 
\begin{enumerate}
\item $A'=\{user, ISPA, ISPB\}$, and
\item $\mathcal{TB}'=\{\textit{transport packets},  \ldots \}$
\end{enumerate}
we derive an example of choosing a supplier for e.g.\ packet transport,
power/elec\-tri\-ci\-ty etc. Promises are then exclusive if ISPA and ISPB are
competitors, for instance.

\section{Conclusion}
We have provided the outline of a process algebra based framework for promise theory.
Using this algebra in combination with conditional guards one can formalize|as other approaches do|how promises might be used
in coming to an agreement. However, in contrast to the static approaches to promise theory mentioned in the introduction,
in the here chosen framework|the algebra of communicating processes ACP| the interaction of promises and the resolution of conflicts can be modelled in a dynamic way.

This formalization is treating promises at a meta-level. There are also
underlying events or processes that the promises suppress|we do not talk about
how the promises are kept, or comment on their reliability; that is a different
matter. Thus our description is at a \emph{promise management level}. At that
level we could say it describes an autonomous process.
\section*{Acknowledgements}
Jan Bergstra and Mark Burgess acknowledge helpful discussions with J\"urgen
Sch\"onw\"alder, School of Engineering and Science, during a working visit  of one week to the Jacobs University
Bremen in October 2006.

 \end{document}